\documentclass[12pt]{article}

 \usepackage[utf8]{inputenc}     
\usepackage[T1]{fontenc}
\usepackage{lmodern}
\usepackage{graphicx}
\usepackage[english]{babel}
\usepackage{amsmath, amsfonts, amssymb} 
\usepackage{tikz}
\usepackage{xcolor}

\newcommand{\llangle}{\langle\!\langle}
\newcommand{\rrangle}{\rangle\!\rangle}
\newcommand{\steady}{|{\cal S}\rangle}

\newcommand{\bea}{\begin{eqnarray}}
\newcommand{\eea}{\end{eqnarray}}
\newcommand{\beano}{\begin{eqnarray*}}
\newcommand{\eeano}{\end{eqnarray*}}
\newcommand{\beq}{\begin{equation}}
\newcommand{\eeq}{\end{equation}}

        \def\cC{{\cal C}}

\def\cS{{\cal S}}

\def\fc{{\mathfrak c}}

\newcommand{\BB}{{\mathbb B}}
\newcommand{\CC}{{\mathbb C}}

\newcommand{\II}{{\mathbb I}}

\newcommand{\cw}{{\mathsf w}}

\newcommand{\mb}[1]{\quad\mbox{#1}\quad}

\begin{document}
\setcounter{page}{1}

\begin{flushright}
LAPTH-Conf-043/16
\end{flushright}

\vspace{20pt}

\begin{center}
\begin{Large}
{\bf  Integrability in out-of-equilibrium systems}
\end{Large}
\\[1.2ex]
\begin{large}
E.~Ragoucy\footnote{eric.ragoucy@lapth.cnrs.fr}
\end{large}

 \vspace{10mm}

 {\it Laboratoire de Physique Th\'eorique LAPTH, CNRS and USBM,\\
BP 110, 74941 Annecy-le-Vieux Cedex, France}

\end{center}

\vspace{1cm}

\begin{abstract}
This short note presents a summary of the articles
\texttt{arXiv:1408.5357}, \texttt{arXiv:1412.5939}, \texttt{arXiv:1603.06796}, \texttt{arXiv:1606.01018},
 \texttt{arXiv:1606.08148} that were done in collaboration 
 with N. CRAMPE, M. EVANS, C. FINN, K. MALLICK and M. VANICAT.
 It presents an approach to the matrix ansatz of out-of-equilibrium systems with boundaries, within the framework of integrable systems, and was presented at ISQS24 (Prague, 2016). \hfill 
\end{abstract}

\section{General context}
Statistical thermodynamics relies on the existence of a thermodynamical equilibrium stationary state, where the probability of a  configuration is given by the Boltzmann distribution:
\begin{equation}
P_{eq}(\mathcal{C}) \sim e^{-\frac{E(\mathcal{C})}{k_B T}}.
\end{equation}
The existence of this distribution allows one to compute many physical data and makes contact with the 'usual' thermodynamical laws. However, the existence of the thermodynamical equilibrium stationary state is quite restrictive, since it implies that there is no particle or energy flow. Then, simple situations such as a metal stick in between two reservoirs with different temperatures cannot be studied within the framework of thermodynamical equilibrium. Nowadays, to generalize this framework, people are considering non-equilibrium stationary states, where the state under consideration does not envolve in time but allows particle or energy currents. Unfortunately, 
the configuration probability $P_{stat}(\mathcal{C})$ of such non-equilibrium stationary state is not known in general.
However for some models, the \textbf{matrix product ansatz} allows exact computations to get $P_{stat}(\mathcal{C})$.
The goal of our work is to understand this matrix ansatz (when it exists) within the integrable system framework and to generalize it to other integrable models, where the matrix ansatz was not known up to now.

To present this 'integrable approach to matrix ansatz', we will first work with one of the simplest (though paradigmatic) model, the TASEP model with open boundaries (see definition below). We will introduce our general approach for this simple model (section \ref{sect:TASEP}), and then present it in its full generality (section \ref{sect:MA}). The general approach applies obviously to more models, but with more generators than the ones used in the TASEP model: we illustrate it on another (reaction-diffusion) model, that we solve using our technics (section \ref{sect:DISSEP}). However, although rather general, our technics still lacks from general principles to fix the form of the matrix ansatz: we show it on a third example, based on multi-species TASEP (section \ref{sect:2TASEP}). Finally, we conclude in section \ref{conclu}. An appendix present to some technical results.

We would like to stress that this short note does not intend to be exhaustive, and many works on the subject should be cited. We refer the interested reader to our original articles, or to reviews such as \cite{DerrReview,BE,Krug2,CKZ,bertin} for more references.

\section{TASEP and Matrix Ansatz\label{sect:TASEP}}

\subsection{The TASEP model}

\begin{figure}[htb]
\begin{center}
 \begin{tikzpicture}[scale=0.7]
\draw (-2,0) -- (12,0) ;
\foreach \i in {-2,-1,...,12}
{\draw (\i,0) -- (\i,0.4) ;}
\draw[->,thick] (-2.4,0.9) arc (180:0:0.4) ; \node at (-2.,1.8) [] {{$\alpha$}};
\draw  (1.5,0.5) circle (0.3) [fill,circle] {};
\draw  (4.5,0.5) circle (0.3) [fill,circle] {};
\draw  (5.5,0.5) circle (0.3) [fill,circle] {};
\draw  (11.5,0.5) circle (0.3) [fill,circle] {};
\draw[->,thick] (1.6,1) arc (180:0:0.4); \node at (2.,1.8) [] {{$1$}};
\draw[->,thick] (4.6,1) arc (180:0:0.4); \draw[thick] (4.8,1.2) -- (5.2,1.6) ;\draw[thick] (4.8,1.6) -- (5.2,1.2) ;
\draw[->,thick] (11.6,1) arc (180:0:0.4) ; \node at (12.,1.8) [] {{$\beta$}};
 \end{tikzpicture}
 \end{center}
\caption{The different transition rates in the TASEP model\label{fig:TASEP}}
\end{figure}
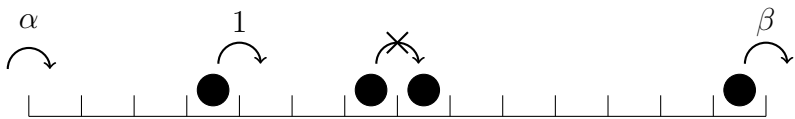

The acronym TASEP stands Totally Asymmetric Simple Exclusion Process, where 'totally asymmetric' indicates that particles are flowing in one direction only, 'simple' means that the jump can be only one step forwards, and 'exclusion' that two particles cannot be simultaneously at the same site. We thus have a one-dimenional lattice on which 'particles' are flowing in one direction only. 
We will consider the case of open boundaries, meaning that the lattice is connected to two infinite dimensional reservoirs, from which the particle enters the lattice (on the left) and leave it (on the right).
The different possible rates of transition are show in figure \ref{fig:TASEP}. A rate $M(\cC,\cC')$ corresponds to the probablity $M(\cC,\cC')\,dt$ that the system jumps from the configuration $\cC'$ to the configuration $\cC$ during a small interval of time $dt$. 

The TASEP is an out-of-equilibrium system, meaning that there exists a stationnary state, although there is a particle current (from the left to the right on figure \ref{fig:TASEP}). As we shall see there is a matrix ansatz that describes this stationnary state. Moreover 
the model is integrable, which makes it interesting for our study.

\subsection{Markov matrix and master equation}
To construct the stationary state, we need to formalize the presentation of the model.
We denote  a configuration of the system by $\mathcal{C}=(\tau_1, \tau_2, \dots, \tau_L)$, with $\tau_j=0$ if the site $j$ is empty and 
$\tau_j=1$ if a particle stands at  site $j$. As an example, we have:
$$
 \begin{tikzpicture}[scale=0.7]
\draw (-1,0) -- (5,0) ;
\foreach \i in {-1,-0.5,...,5}
{\draw (\i,0) -- (\i,0.2) ;}
\draw  (-0.25,0.25) circle (0.15) [fill,circle] {};
\draw  (1.25,0.25) circle (0.15) [fill,circle] {};
\draw  (1.75,0.25) circle (0.15) [fill,circle] {};
\draw  (4.75,0.25) circle (0.15) [fill,circle] {};
\draw  (3.25,0.25) circle (0.15) [fill,circle] {};
\draw[->,thick] (5.5,0.25) -- (6.5,0.25);
\node at (10.5,0.2) [] {$(0,1,0,0,1,1,0,0,1,0,0,1).$};
 \end{tikzpicture}
$$

With this notation, the rates presented in figure \ref{fig:TASEP} can be recasted as
\begin{itemize}
 \item {In the bulk,} particles can jump to the right at the rate: $M(\{....,0,1,....\};\{....,1,0,...\})=1$
 \item {On the left boundary,} particles enter at the rate: $M(\{1,....\};\{0,...\})=\alpha$
\item {On the right boundary,} particles leave at the rate: $M(\{....,0\};\{....,1\})=\beta$
\end{itemize}
where dots stand for arbitrary values of the $\tau_j$'s.

We denote by $P_t(\mathcal{C})$ the probability to be in configuration $\mathcal{C}$ at time $t$.
The time evolution of this probibility is governed by the different rates that define the model. In fact, 
performing a balance among the possible transitions between different configurations, we get
\begin{equation}
P_{t+dt}(\mathcal{C})=\sum_{\mathcal{C}'\neq \mathcal{C}}M(\mathcal{C},\mathcal{C}')dt\,P_t(\mathcal{C}')
 +\Big(1-\sum_{\mathcal{C}'\neq \mathcal{C}}M(\mathcal{C}',\mathcal{C})dt\,\Big) P_t(\mathcal{C}).
\end{equation}
In words, it just states that the probalility to be in configuration $\cC$ at time $t+dt$ is the sum of two terms:
$(i)$ the probability to 
jump in this configuration starting from another one between $t$ and $t+dt$,  $(ii)$ 
the probability to stay in this configuration between $t$ and $t+dt$.

For $dt\to0$ this equation can be recasted as
 \begin{equation}\label{eq:master}
 \frac{d P_t(\mathcal{C})}{dt}=\sum_{\mathcal{C}'\neq \mathcal{C}}M(\mathcal{C},\mathcal{C}')P_t(\mathcal{C}')
 -\sum_{\mathcal{C}'\neq \mathcal{C}}M(\mathcal{C}',\mathcal{C}) P_t(\mathcal{C}).
\end{equation}
Equation \eqref{eq:master} is called the 
Master equation.

Since at each site, we have 2 possibilities (0 or 1, i.e. empty or occupied),  there is a $\CC^2$ "local" space to describe the different probabilities. Then, for
 $L$ sites we get a $\big(\CC^2\big)^{\otimes L}$ total space, so that we can
  gather all the probabilities in a vector
\begin{equation}
|P_t\rangle=\left(
\begin{array}{c}
P_t(\ (0,\dots,0,0,0)\ )\\
P_t(\ (0,\dots,0,0,1)\ )\\[1ex]
P_t(\ (0,\dots,0,1,0)\ )\\
P_t(\ (0,\dots,0,1,1)\ )\\[1ex]
\vdots\\
P_t(\ (1,\dots,1,1,1)\ )
\end{array}
\right)\ \in\quad \big(\CC^2\big)^{\otimes L}
\end{equation}
and the Master equation takes a vectorial form
 \begin{equation}
\frac{d |P_t\rangle}{dt}=M\ |P_t\rangle\ \in\ \big(\CC^2\big)^{\otimes L},
\end{equation}
where the Markov matrix $M$ can be written in terms of local jump operators
\begin{equation}\label{eq:markov}
 M =  {B_1} + {\sum_{\ell=1}^{L-1} m_{\ell,\ell+1}} + {\overline{B}_L}\ \in\ \big(End(\CC^2)\big)^{\otimes L}.
\end{equation}
In \eqref{eq:markov} we have used the standard auxiliary space notation for tensor products of $\CC^2$ spaces, i.e. the indices indicate on 
which copies of $\CC^2$ in the total space $\big(\CC^2\big)^{\otimes L}$, the matrices act non-trivially:
\begin{eqnarray}
&&m_{\ell,\ell+1}=\underbrace{\II\otimes\cdots\otimes\II}_{\ell-1}\otimes \underbrace{m}_{\ell,\ell+1}\otimes
\underbrace{\II\otimes\cdots\otimes\II}_{L-1-\ell}\ ;\\
&& B_1=\underbrace{ B}_1\otimes\underbrace{\II\otimes\cdots\otimes\II}_{L-1} \quad
; \quad \overline B_L=\underbrace{\II\otimes\cdots\otimes\II}_{L-1}\otimes \underbrace{\overline B}_L
\end{eqnarray}
where $\II$ is the $2\times2$ identity matrix.

The local matrices $m$, $B$ and $\overline B$ corresponding to the rates described above take the following form:
\begin{eqnarray}\label{eq:local}
&&\underbrace{B =\left( \begin {array}{cc} 
-\alpha&0\\ 
\alpha&0
\end {array} \right)}_{\in End(\CC^2)}
\ ;\ 
\underbrace{m=\left( \begin {array}{cccc} 
0&0&0&0\\ 
0&0&1&0\\
0&0&-1&0\\
0&0&0&0
\end {array} \right)}_{\in End(\CC^2\otimes\CC^2)}
\ ;\ 
\underbrace{\overline{B} =\left( \begin {array}{cc} 
0&\beta\\ 
0&-\beta
\end {array} \right)}_{\in End(\CC^2)}.
\end{eqnarray}

We remind that our goal is to compute the probability distribution corresponding to the (time independent) stationary state, $\cS(\cC)$, which in the vectorial form amounts to solve the zero-eigenvalue problem for the Markov matrix \eqref{eq:markov}: 
$$M\steady=0.$$

\subsection{Matrix ansatz}
The matrix ansatz was introduced in \cite{DEHP}, and then developped by many authors for different models. 
In this formalism, to get the probability of a given configuration $\cC$ in the stationary state, one first associates 
to each empty site an operator $E$, and to each occupied site an operator $D$, forming an algebra (see below). 
Then, one can uniquely associate to the configuration $\cC$ a 'word' $\cw(\cC)$ of length $L$ in $E$ and $D$. 
One also considers two vectors $\llangle W|$ and $|V\rrangle$ belonging to two additional\footnote{These spaces 
are usually called auxiliary spaces by the statistical physicists, but they are different from the auxiliary spaces used 
in integrable systems, so that we choose to call them additional.} spaces that form two representations of the 
$(E,D)$ algebra. Let us stress that the additional spaces have nothing to do with the space $(\CC^2)^{\otimes L}$ introduced before. In particular, the vectors $\llangle W|$ and $|V\rrangle$ are scalar with respect to the $(\CC^2)^{\otimes L}$ space, and commute with the matrices $m$, $B$ and $\overline B$.

Now, the matrix ansatz states that the probability to be in the configuration $\cC$ in the stationary state is given by
\begin{equation}
\cS(\cC) = \frac{\llangle W|\cw(\cC)|V\rrangle}{Z_L}
\mb{with} Z_L=\llangle W| (D+E)^L|V\rrangle,
\end{equation}
provided $E$ and $D$ obey the following algebraic relations (for the TASEP model)
\begin{equation}\label{DE:tasep}
DE = D+E 
\end{equation}
and the vectors $\llangle W|$ and $|V\rrangle$ are such that
\begin{equation}\label{WV:tasep}
D |V \rrangle = \frac{1}{\beta}|V\rrangle \quad;\quad
\llangle W| E = \frac{1}{\alpha}\llangle W|.
\end{equation}
Let us stress that not only there is a proof that the matrix ansatz provides the right weights for the probabilities (see below where the proof is reminded), but also it gives an explicit and recursive way to explicitly compute them. We illustrate it on an example.
\subsubsection*{Example}
If one consider a model with $L=5$ sites and a configuration $\cC=(0,1,0,1,1)$, one has
$\cw(0,1,0,1,1)=EDEDD$, so that
$$
{\cal S}( \begin{tikzpicture}[scale=0.7]
\draw (4,1) -- (6.5,1) ;
\foreach \i in {4,4.5,...,6.5}
{\draw (\i,1) -- (\i,1.2) ;}
\draw  (4.75,1.25) circle (0.15) [fill,circle] {};
\draw  (5.75,1.25) circle (0.15) [fill,circle] {};
\draw  (6.25,1.25) circle (0.15) [fill,circle] {};
 \end{tikzpicture}
 \displaystyle{)=\frac{\llangle W|EDEDD|V\rrangle}{Z_5}}
\mb{with} Z_5=\llangle W| (D+E)^5|V\rrangle.
$$
Using first relation \eqref{DE:tasep} on the product $DE$ and then relations \eqref{WV:tasep} for $E$ on the left and 
for $DD$ on the right one  easily computes that
$$
\llangle W|EDEDD|V\rrangle = \llangle W|E(D+E)DD|V\rrangle=
\frac{1}{\alpha\,\beta^2}\,\llangle W|(D+E)|V\rrangle=\frac{\alpha+\beta}{\alpha^2\,\beta^3}\,\llangle W|V\rrangle.
$$
In the same way, one gets
\begin{eqnarray*}
Z_5 &=& \Big\{
\beta^5 + \alpha \beta^4 (1 + 4 \beta) + 
 \alpha^2 \beta^3 (1 + 4 \beta + 9 \beta^2) + 
 \alpha^3 \beta^2 (1 + 4 \beta + 9 \beta^2 + 14 \beta^3) + 
 \\
&& 
+ \alpha^4 \beta (1 + 4 \beta + 9 \beta^2 + 14 \beta^3 + 14 \beta^4)
+ \alpha^5 (1 + 4 \beta + 9 \beta^2 + 14 \beta^3 + 14 \beta^4) 
\Big\} \frac{\llangle W|V\rrangle}{\alpha^5\beta^5}
\end{eqnarray*}
Then, up to the proof that there exist vectors $\llangle W|$ and $|V\rrangle$ such that $\llangle W|V\rrangle\neq 0$, one gets an explicit expression for $
{\cal S}( \begin{tikzpicture}[scale=0.7]
\draw (4,1) -- (6.5,1) ;
\foreach \i in {4,4.5,...,6.5}
{\draw (\i,1) -- (\i,1.2) ;}
\draw  (4.75,1.25) circle (0.15) [fill,circle] {};
\draw  (5.75,1.25) circle (0.15) [fill,circle] {};
\draw  (6.25,1.25) circle (0.15) [fill,circle] {};
 \end{tikzpicture}
 \displaystyle{)}.$
The construction of such vectors is also known for the TASEP, see \cite{DEHP} for more details.

\subsection{The matrix ansatz gives the right weights of the stationary state}
We reproduce  this proof, not in its original version, but using an integrable system formalism that will be adapted to our generalisation.
 The vector $\steady$ can be written in the compact form
\begin{equation}
 \steady = \frac{1}{Z_L}{\llangle W|}\, A_{{1}}\, \,A_{{2}}  \cdots 
  \,A_{{L}}\, {|V\rrangle }\,\in\ (\CC^2)^{\otimes L}
  \mb{with} A=\left(\begin{array}{c}E\\D\end{array}\right)
\end{equation}
where we have again used the auxiliary space notation for tensor products of $\CC^2$ spaces, meaning that
$$
A_\ell = \underbrace{\II\,\otimes \cdots  \otimes \,\II}_{\ell-1}\otimes A\otimes 
\underbrace{\II\,\otimes \cdots  \otimes \,\II}_{L-\ell}\,,\quad\forall \ell
\mb{so that}
A_{{1}}\, \,A_{{2}}  \cdots 
  \,A_{{L}}=\underbrace{A\,\otimes \,A \otimes \cdots  \otimes \,A}_L.
$$

Using the form of the local markov matrix $m$, see \eqref{eq:local},
one can show that the {bulk relation} $DE=D+E$ is equivalent to
 \begin{eqnarray}\label{eq:MA1}
&& {m}\ A \otimes A = {\bar A} \otimes  A-A \otimes {\bar A}
 \mb{with}
\bar A = \left(\begin{array}{c}1\\-1
 \end{array}\right).
\end{eqnarray}
Then
\beano
{\Big(m_{12}+m_{23}+\cdots+m_{L-1,L}\Big)}\ 
A_{1}\, \,A_{2}  \cdots  \,A_{L}
&=&
 \Big({\bar A_{1}}\, \,A_{2} -A_{1}\,  {\bar A_{2}}\Big) A_{3} \,\cdots\, A_{L} 
 \\
&&+ A_{1}\Big( {\bar A_{2}}\, A_{3}  
- A_{2}\, {\bar A_{3}} \Big) A_{4} \,\cdots \, A_{L} 
 \\ && \qquad\qquad\qquad\vdots 
 \\
 &&+ A_{1}\, \cdots \, A_{L-2} \Big({\bar A_{L-1}} \, A_{L}
- A_{L-1} \, {\bar A_{L}}\Big) 
\eeano
where the first line corresponds to the action of $m_{12}$, the second line to the action of $m_{23}$, ..., up to the action of $m_{L-1,1}$. One sees that we get a telescoping sum, so that finally only the first and the last terms remain:
\beano
{\Big(m_{12}+m_{23}+\cdots+m_{L-1,L}\Big)}\ 
A_{1}\, \,A_{2}  \cdots  \,A_{L}
&=&
 {\bar A_{1}}\, \,A_{2}  \cdots  \,A_{L}
-A_{1}\,  \cdots \,A_{L-1}  \,{\bar A_{L}}
\eeano
 This equality holds in the algebra, without the use of the vectors $\llangle W|$ and $|V\rrangle$. Obviously it is still valid when applied to these vectors.

One can also show that the left (resp. right)
 boundary condition
${\llangle W|E=\frac{1}{\alpha}\llangle W|}$ (resp. ${D|V\rrangle=\frac{1}{\beta}|V\rrangle}$) are equivalent to 
 \begin{eqnarray}\label{eq:MA2}
&&\displaystyle \llangle W|B\,A = -\llangle W|{\bar A} 
\mb{and} \bar B\,A|V\rrangle = {\bar A}|V\rrangle 
\end{eqnarray}
with again $ A=\left(\begin{array}{c}E \\ D\end{array}\right)$ and
${\bar A = \left(\begin{array}{c}1\\-1\end{array}\right)}$.
The relations \eqref{eq:MA2} are recasted as 
\begin{eqnarray*}
&& \llangle W|B_1\,A_{1}\, \,A_{2}  \cdots  \,A_{L} = -\llangle W|\bar A_{1}\, \,A_{2}  \cdots  \,A_{L}
\\
&&\bar B_L\,A_{1}\, \,A_{2}  \cdots  \,A_{L}|V\rrangle = A_{1}\, \,A_{2}  \cdots  A_{{L-1}}\,\bar A_{L}|V\rrangle .
\end{eqnarray*}

Putting together these algebraic relations we get
\begin{eqnarray*}
M\llangle W|\, A_{{1}}\, \,A_{{2}}  \cdots  \,A_{{L}}\, |V\rrangle &=&
\left( {\sum_{\ell=1}^{L-1} m_{\ell,\ell+1}} +B_1+ {\overline{B}_L}\right)
\llangle W|\, A_{{1}}\, \,A_{{2}}  \cdots \,A_{{L}}\, |V\rrangle
=0.
\end{eqnarray*}
This shows that the steady state $\steady$ is proportional to 
$\llangle W|\, A_{{1}}\, \,A_{{2}}  \cdots \,A_{{L}}\, |V\rrangle$.
One fixes the normalisation by demanding that all the probabilities add up to 1, so that
\begin{eqnarray*}
\steady &=& \frac{1}{Z_L}{\llangle W|}\, A_{{1}}\, \,A_{{2}}  \cdots \,A_{{L}}\, {|V\rrangle }
=  \frac{1}{Z_L} {\llangle W|}\, A\,\otimes \,A \otimes \cdots  \otimes \,A\, {|V\rrangle } 
\end{eqnarray*}
where we remind that $Z_L= \llangle W| (D+E)^L|V\rrangle$.

\subsection{Integrability of the TASEP model}
As already mentioned, the model is integrable, and we briefly summarize the basic ingredients that are related to its integrability. 
\subsubsection{In the bulk: the $R$-matrix}

\beano
&& 
R(x)=\left( \begin {array}{cccc} 
1&0&0&0\\ 0&0&x&0\\ 0&1&1-x&0\\ 0&0&0&1
\end {array} \right)\ \in\, End(\CC^2\otimes\CC^2) 
\eeano
where $x\in\,\CC$ is the spectral parameter. It is related to the local matrix $m$ in the usual way
 $$P \left.\frac{d}{dx}R(x)\right|_{x=1}=-m \mb{with} P=\left( \begin {array}{cccc} 
1&0&0&0\\ 0&0&1&0\\ 0&1&0&0\\ 0&0&0&1\end {array} \right).$$

Integrability in the bulk is ensured by the {Yang-Baxter} equation
\begin{eqnarray*}
&& R_{12}\!\left(\frac{x_1}{x_2}\right) R_{13}\!\left(\frac{x_1}{x_3}\right) R_{23}\!\left(\frac{x_2}{x_3}\right) =
 R_{23}\!\left(\frac{x_2}{x_3}\right) R_{13}\!\left(\frac{x_1}{x_3}\right) R_{12}\!\left(\frac{x_1}{x_2}\right)
\end{eqnarray*}
where we used again the auxiliary space notation, e.g. $R_{12}(x)=R(x)\otimes\II$ and 
$R_{23}(x)=\II\otimes R(x)$.
The $R$-matrix  is {unitary} and regular: 
$$R_{12}(x)R_{21}(1/x)=1 \mb{and}
R(1)=P.$$

\subsubsection{{On the boundaries: $K$-matrices}}

 \begin{equation} 
K(x)=\left( \begin {array}{cc} 
\displaystyle{\frac{(-x \alpha+\alpha-1)x}{x \alpha-x-\alpha}}&0\\[2ex]
\displaystyle{\frac{\alpha(x^2-1)}{x\alpha-x-\alpha}}&1
\end {array} \right),\quad
\bar K(x)= \left( \begin {array}{cc} 
1&\displaystyle{-\frac{(x^2-1)\beta}{-x^2\beta+x\beta-x}}\\[2ex]
0&\displaystyle{\frac{x\beta-x-\beta}{-x^2\beta+x\beta-x}}
\end {array} \right)
\end{equation}
They are connected to the local matrices $B$ and $\overline B$:
$$\left.\frac{d}{dx}K(x)\right|_{x=1}=-2B \mb{and} \left.\frac{d}{dx}\bar K(x)\right|_{x=1}=2\bar B.$$

They satisfy the {reflection equation}
\begin{equation}
 R_{12}\left(\frac{x_1}{x_2}\right) K_1(x_1) R_{21}(x_1 x_2) K_2(x_2)=
 K_2(x_2)R_{12}(x_1 x_2)K_1(x_1)R_{21}\left(\frac{x_1}{x_2}\right)
\end{equation}
and are {unitarity} and {regular} 
$$K(x)K(1/x)=1\mb{and}K(1)=1.$$

\subsubsection{{Integrability:}}

 We define the usual double row {transfer matrix} \cite{sklyanin}

 \begin{equation}\label{eq:doublerow}
{t(x)}= tr_0\Big( \widetilde K_0(x)\ R_{0L}(x)\cdots R_{01}(x)\ K_0(x)\ R_{10}(x)\cdots R_{L0}(x)  \Big),
\end{equation}
where $\widetilde K$ is linked to $\bar K$ in the following way
\begin{equation}
 \bar K_1(x)= tr_0\Big( \widetilde{K}_0(\frac1x)R_{01}(\frac1{x^2})P_{01} \Big).
\end{equation}
It generates the Markov matrix
  \begin{equation}
-\frac{1}{2} \left. \frac{dt(x)}{dx}\right|_{x=1}=-\frac{1}{2} K_1'(1)-\sum_{k=1}^{L-1}P_{k,k+1}R'_{k,k+1}(1)
+\frac{1}{2}\bar K_L'(1)=M
\end{equation}
and integrability of the model is ensured by the commutativity of the transfer matrix
 $[t(x),t(x')]=0$. This last relation is proven in the usual way \cite{sklyanin}.

 {Integrability} ensures that we can construct exactly the eigenvectors of the transfer matrix (and thus of the Markov matrix) using for instance the Bethe ansatz. 
The Bethe vectors are eigenvector of the transfer matrix $t(x)\,\BB(\bar u) = \lambda(x,\bar u)\,\BB(\bar u)$ when the set of parameters $\bar u$ obeys the so-called Bethe equations.
However the {stationary state} is not easily obtained in this way since one should solve the Bethe equations to identify it.
Our goal is to recover the matrix ansatz solution from R and K matrices.

\subsection{Matrix ansatz relations from R and K matrices}

Let us define the vector 
\begin{equation}
 A(x)=  \left(\begin{array}{c}E-1+x\\D-1+\frac{1}{x}
 \end{array}\right)\quad \Rightarrow\quad A(1)=  \left(\begin{array}{c}E\\D
 \end{array}\right)\equiv A, \ A'(1)=  \left(\begin{array}{c}1\\-1
 \end{array}\right)\equiv \bar A.
\end{equation}
We impose on the vector $A(x)$ the following well-known relations.

\textbf{Zamolodchikov-Faddeev (ZF) relation \cite{ZF}}:
\begin{equation}\label{eq:ZF}
 R_{12}\!\left(\frac{x}{y}\right)A_1(x)\,A_2(y)=A_2(y)\,A_1(x).
\end{equation}

\textbf{Ghoshal-Zamolodchikov (GZ) relations \cite{GZ}:}
\begin{eqnarray}\label{eq:GZ}
{\llangle W|A(x)=\llangle W|K(x)A\left(\frac{1}{x}\right)} \mb{and}
{A(x)|V\rrangle = \bar K(x)A\left(\frac{1}{x}\right)|V\rrangle}.
\end{eqnarray}

It turns out that these relations, although they depend on spectral parameters simplify drastically (due to the explicit expressions of $R(x)$, $K(x)$, $\overline K(x)$ and $A(x)$), and are equivalent to the matrix ansatz relations
\begin{equation}
DE=D+E\,,\quad \llangle W| E= \frac{1}{\alpha}\llangle W|\,,\quad D |V\rrangle = \frac{1}{\beta}|V\rrangle.
\end{equation}

Moreover, taking the derivative with respect to $x$ and setting $x=y=1$ in \eqref{eq:ZF} and \eqref{eq:GZ}, one obtains exactly the relations \eqref{eq:MA1} and \eqref{eq:MA2} that are needed for the matrix ansatz.
This suggest the following general "integrable approach" to the matrix ansatz.

\section{Integrable approach to matrix ansatz\label{sect:MA}}
We start with an integrable model described by a unitary {$R$-matrix} $R_{12}(x)$ 
of size $r^2\times r^2$, which obeys {YBE}, and the corresponding unitary 
{boundary matrices} 
$K(x)$ and $\bar K(x)$, of size $r\times r$, obeying the {reflection equation.}
Integrability is ensured by the double-row transfer matrix \eqref{eq:doublerow}. 

\subsection{Construction of the matrix ansatz}
We introduce the vector
$
 A(x)=\left(
 \begin{array}{c}
  X_1(x)\\
  X_2(x)\\
   \vdots\\
  X_r(x)
 \end{array}
\right)$
where the elements $X_1(x),\dots,X_r(x)$ belong to some algebra, defined by the following relations

{\textbf{ZF relations (in the bulk):}}
 \begin{equation}
  R_{12}\left(\frac{x}{y}\right) A_1(x)A_2(y)=A_2(y)A_1(x)
 \end{equation}
The associativity of this algebra ensured by the Yang-Baxter equation.
  Another consistency relation for this algebra is ensured by the unitarity of the R-matrix.
 
The ZF relation implies (through derivation w.r.t. $x$ and setting $x=y=1$)
\begin{equation}
m\ A_1(1)\,A_2(1)=A'_1(1)\,{A_2(1)}-{A_1(1)}\,A'_2(1)
\mb{with} m=-  {P R_{12}'(1)}.
 \end{equation}
 
N.B.: Nothing in this construction implies that $\bar A=A'(1)$ is necessarily scalar, despite it was scalar  for the TASEP model. We illustrate it on an example in section \ref{sect:DISSEP}.

{\textbf{GZ relation (on the boundaries):}}
\begin{equation}
   \llangle W|A(x)=\llangle W|K(x)A\left(\frac{1}{x}\right) \mb{and} A(x)|V\rrangle=\bar K(x)A\left(\frac{1}{x}\right)|V\rrangle
\end{equation}
Again the consistency of these relations is ensured by the reflection equation and the unitarity of the K matrices.
  
They imply (through derivation w.r.t. $x$ and setting $x=1$)
\begin{eqnarray}
&&\llangle W|\,BA(1)=-\llangle W|{A'(1)}\mb{with} B=-\frac12 K'(1),\\
&&  {\bar B}A(1)|V\rrangle = {A'(1)}|V\rrangle \mb{with} \bar B=\frac12 \bar K'(1).
\end{eqnarray}

Then the vector 
\begin{equation}
\steady= \frac{1}{Z_L}\,\llangle W|A_1(1)A_2(1)\cdots A_L(1)|V\rrangle,
\end{equation}
 
is the stationary state of the process (we have again a telescopic sum):

\begin{equation}
M\steady=0
\mb{for} M={B_1} + {\sum_{\ell=1}^{L-1} m_{\ell,\ell+1}} + {\overline{B}_L}.
\end{equation}

\subsection{Inclusion of inhomogeneities}
This approach can be further generalized.
We define the vector 
\begin{equation}
|{\cal S}(\theta_1,\dots,\theta_L)\rangle=\llangle W| A_1(\theta_1)\cdots A_L(\theta_L) |V\rrangle.
\end{equation}

Starting from $|{\cal S}(\theta_1,\dots,\theta_L)\rangle$ one chooses a vector $A_i(\theta_i)$ for some $i$. We move it to the right using the ZF algebra, bounce it on $|V\rrangle$ thanks to GZ relation, then move it to the left (again with the ZF algebra) and bounce it on $\llangle W|$ (with the GZ relation) before moving it back to its original place using the ZF algebra. 
Gathering all the $R$ and $K$ matrices one gets through this procedure, one shows that\footnote{The second relation is obtained in the same way, but moving to the left first.}
\begin{eqnarray*}
&&  |{\cal S}(\theta_1,\dots,\theta_L)\rangle = t(\theta_i | \bar \theta ) \ |{\cal S}(\theta_1,\dots,\theta_L)\rangle,
\\
&& |{\cal S}(\theta_1,\dots,\theta_L)\rangle = t(1/\theta_i | \bar \theta ) \ |{\cal S}(\theta_1,\dots,\theta_L)\rangle,
 \end{eqnarray*}
with the {inhomogeneous} transfer matrix
\begin{equation}
 t(x | {\bar \theta})=tr_0\Big(\ \widetilde K_0(x)\ R_{0L}(\frac{x}{{\theta_L}})\cdots R_{01}(\frac{x}{{\theta_1}})\ K_0(x)\ R_{10}(x{\theta_1})\cdots R_{L0}(x{\theta_L})  \ \Big).
\end{equation}

If one assumes moreover a {crossing symmetry} for the R and K matrices, 
one gets a crossing relation on the transfer matrix:
 \begin{equation} 
 t(x|\bar \theta)=
 \Big(\lambda(x | \bar \theta)-1\Big)\; t(1/xq | \bar \theta)
\end{equation}
where $\lambda(x | \bar \theta)$ is a function that depends on the model.
Then, one can perform an interpolation in $x$ to obtain
\begin{equation}
   t(x | \bar \theta ) \ |{\cal S}(\theta_1,\dots,\theta_L)\rangle =\lambda(x |\bar \theta)|{\cal S}(\theta_1,\dots,\theta_L).
\end{equation}
This shows that the 'generalized steady state' is an eigenvector of the transfer matrix $t(x | \bar \theta )$.
Remark that, contrarily to the Bethe ansatz, there is no Bethe parameters (and no Bethe eqs).
However, we get only one state. 


\section{Example: a reaction-diffusion model  (DiSSEP)\label{sect:DISSEP}}
We illustrate our method on a new model that we constructed from $R$ and $K$ matrices \cite{progress}. 
\subsection{Description of the model, R and K matrices}
We start by introducing the following matrices.
\begin{equation}
 R(x)=\left( \begin {array}{cccc} 
\frac{\kappa(x+1)}{\kappa(x+1)+x-1}&0&0&\frac{x-1}{\kappa(x+1)+x-1}\\ 
0&\frac{\kappa(x-1)}{\kappa(x-1)+x+1}&\frac{x+1}{\kappa(x-1)+x+1}&0\\
0&\frac{x+1}{\kappa(x-1)+x+1}&\frac{\kappa(x-1)}{\kappa(x-1)+x+1}&0\\
\frac{x-1}{\kappa(x+1)+x-1}&0&0&\frac{\kappa(x+1)}{\kappa(x+1)+x-1}
\end {array} \right).
\end{equation}
This $R$-matrix satisfies the Yang-Baxter equation. It is unitary and regular.

 \begin{equation}
K(x)=\left( \begin {array}{cc} 
\frac{(x^2+1)((x^2-1)(\gamma-\alpha)+4x\kappa)}{2x((x^2-1)(\alpha+\gamma)+2\kappa(x^2+1))}&\frac{(x^2-1)((x^2+1)(\gamma-\alpha)+2x(\alpha+\gamma))}{2x((x^2-1)(\alpha+\gamma)+2\kappa(x^2+1))}\\[2ex] 
-\frac{(x^2-1)((x^2+1)(\gamma-\alpha)-2x(\alpha+\gamma))}{2x((x^2-1)(\alpha+\gamma)+2\kappa(x^2+1))}&-\frac{(x^2+1)((x^2-1)(\gamma-\alpha)-4x\kappa)}{2x((x^2-1)(\alpha+\gamma)+2\kappa(x^2+1))}
\end {array} \right),
\end{equation}
\begin{equation}
\bar K(x)= \left( \begin {array}{cc} 
\frac{(x^2+1)((x^2-1)(\delta-\beta)+4x\kappa)}{2x(-(x^2-1)(\delta+\beta)+2\kappa(x^2+1))}&\frac{(x^2-1)((x^2+1)(\delta-\beta)-2x(\delta+\beta))}{2x(-(x^2-1)(\delta+\beta)+2\kappa(x^2+1))}\\[2ex] 
-\frac{(x^2-1)((x^2+1)(\delta-\beta)+2x(\delta+\beta))}{2x(-(x^2-1)(\delta+\beta)+2\kappa(x^2+1))}&-\frac{(x^2+1)((x^2-1)(\delta-\beta)-4x\kappa)}{2x(-(x^2-1)(\delta+\beta)+2\kappa(x^2+1))}
\end {array} \right).
\end{equation}
These $K$-matrices satisfy the reflection equation. They are unitary and regular.
 
\begin{figure}[htb]
\begin{center}
 \begin{tikzpicture}[scale=0.7]
\draw (-2,0) -- (12,0) ;
\foreach \i in {-2,-1,...,12}
{\draw (\i,0) -- (\i,0.4) ;}
\draw[->,thick] (-2.4,0.9) arc (180:0:0.4) ; \node at (-2.,1.8) [] {$\alpha$};
\draw[->,thick] (-1.6,-0.1) arc (0:-180:0.4) ; \node at (-2.,-0.8) [] {$\gamma$};
\draw  (1.5,0.5) circle (0.3) [fill,circle] {};
\draw  (4.5,0.5) circle (0.3) [fill,circle] {};
\draw  (5.5,0.5) circle (0.3) [fill,circle] {};
\draw  (8.5,3.1) circle (0.3) [fill,circle] {};
\draw  (9.5,3.1) circle (0.3) [fill,circle] {};
\draw[->,thick] (1.4,1) arc (0:180:0.4); \node at (1.,1.8) [] {$\kappa^2$};
\draw[->,thick] (1.6,1) arc (180:0:0.4); \node at (2.,1.8) [] {$\kappa^2$};
\node at (5,1.1) [rotate=-90] {$\Big{\{}$};
\draw[->,thick] (5,1.3) -- (5,2.8); \node at (5.3,2.2) [] {$1$};
\node at (9,2.5) [rotate=90] {$\Big{\{}$};
\draw[->,thick] (9,2.3) -- (9,0.8); \node at (9.3,2) [] {$1$};
\draw[->,thick] (11.6,1) arc (180:0:0.4) ; \node at (12.,1.8) [] {$\beta$};
\draw[->,thick] (12.4,-0.1) arc (0:-180:0.4) ; \node at (12.,-0.8) [] {$\delta$};
 \end{tikzpicture}
\vspace{-2.4ex}
\end{center}
 \caption{The transition rates in the DiSSEP model\label{fig:rdm}}
\end{figure}
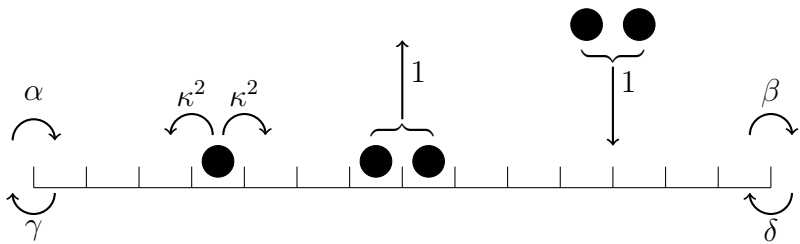

 From these matrices, we get a Markov matrix
$$M={B_1} + {\sum_{\ell=1}^{L-1} m_{\ell,\ell+1}} + {\overline{B}_L}$$
where the local jump operators read
\begin{eqnarray} \label{eq:localw8vertex}
&&B =\kappa\,K'(1)=\left( \begin {array}{cc} 
-\alpha&\gamma\\ 
\alpha&-\gamma
\end {array} \right),\  \
{\overline{B}} =-\kappa\bar K'(1)=\,\left( \begin {array}{cc} 
-\delta&\beta\\ 
\delta&-\beta
\end {array} \right)\\
&&m=2\kappa\,P R'(1)=\left( \begin {array}{cccc} 
-1&0&0&1\\ 
0&-\kappa^2&\kappa^2&0\\
0&\kappa^2&-\kappa^2&0\\
1&0&0&-1
\end {array} \right).
\end{eqnarray}
The corresponding rates of transition are summarized in figure \ref{fig:rdm}. One sees that the model is of SSEP type (i.e. a model where the particles move to the right or to the left with the same rate) but where pairs of particles can condensate or evaporate from the bulk, hence the name DiSSEP, "Di" standing for dissipative.

\subsection{Construction of the algebra needed for the matrix ansatz}

We introduce the vector $A(x)$ 
 \begin{equation}
 A( x)= \left( \begin{array}{c}
G_1 x+G_2+\frac{G_3}{ x}\\[1ex]
-G_1 x+G_2-\frac{G_3}{ x}           
              \end{array} \right)
\end{equation}
and impose the ZF and GZ relations. We obtain

\begin{equation}\label{dissep-ZF}
R_{12}\left(\frac{x_1}{x_2}\right)A_1(x_1)\,A_2(x_2)=A_2(x_2)\,A_1(x_1)\  \Leftrightarrow \ \left\{ \begin{aligned}
          & \phi\ G_1G_2=G_2G_1 \\
          & G_1G_3=G_3G_1 \\
          & \phi\ G_2G_3=G_3G_2
         \end{aligned}
 \right. \mb{with} \phi=\frac{\kappa-1}{\kappa+1}.
\end{equation}

 \begin{eqnarray}\label{dissep-GZ1}
 \llangle W|A(x)&=&\llangle W|K(x)A\left(\frac{1}{x}\right) \ \Leftrightarrow \ \llangle W |\big( G_1-c\,G_2-a\,G_3 \big)=0
 \\
\label{dissep-GZ2}
 A(x)|V\rrangle &=& \bar K(x)A\left(\frac{1}{x}\right)|V\rrangle \ \Leftrightarrow \ \big( G_3-b\,G_1-d\,G_2 \big) | V \rrangle =0
\end{eqnarray}
with $\displaystyle{a=\frac{2\kappa-\alpha-\gamma}{2\kappa+\alpha+\gamma}}$, $\displaystyle{c=\frac{\gamma-\alpha}{2\kappa+\alpha+\gamma}}$, $\displaystyle{b=\frac{2\kappa-\delta-\beta}{2\kappa+\delta+\beta}}$ and $\displaystyle{d=\frac{\beta-\delta}{2\kappa+\delta+\beta}}$.

\subsection{{Matrix ansatz:}}
\begin{equation}
 A(1)= \left( \begin{array}{c} G_2+(G_1+G_3) \\ G_2-(G_1+G_3) \end{array} \right) 
 \mb{and}  
  A'(1)= (G_1-{G_3})\, \left( \begin{array}{c} 1 \\ -1\end{array} \right)
\end{equation}
From $A(1)\equiv A$, we get the expressions $D=G_2-G_1-G_3$ and $E=G_2+G_1+G_3$. Remark that 
$A'(1)\equiv \bar A$ is not a scalar anymore, since it involves the generator $H=G_1-{G_3}$. 
Nevertheless one can still apply the matrix ansatz to get the steady state. Note that the calculations are easier in the $G$ basis, rather than in the $E,D,H$ basis. For instance the
normalization factor reads
\begin{equation}
 Z_L=\llangle W| (D+E)^L|V\rrangle=2^L\llangle W| G_2^L|V\rrangle.
\end{equation}
In fact we are able to compute any word in $G_1$, $G_2$, $G_3$, for more details see \cite{phi}.

\subsection{Calculation of physical observables}
One can compute the densities at one site, defined as ($i=1,2,...,L$):
 \begin{eqnarray*}
 \langle n_i \rangle &=& \frac{\llangle W| (D+E)^{i-1}D(D+E)^{L-i} |V\rrangle}{\llangle W| (D+E)^L |V\rrangle}
=\frac{1}{2}\frac{\llangle W| G_2^{i-1}(G_2-G_1-G_3)G_2^{L-i} |V\rrangle}{\llangle W| G_2^L |V\rrangle}.
\end{eqnarray*}
Using solely the relation \eqref{dissep-ZF}, \eqref{dissep-GZ1} and \eqref{dissep-GZ2}, we get
\begin{equation}\label{eq:densite}
\langle n_i \rangle = 
\frac{1}{2}-\frac{c\phi^{i-1}+ad\phi^{L+i-2} 
 +d\phi^{L-i}+bc\phi^{2L-i-1}}{2(1-ab\phi^{2L-2})}.
\end{equation}
Remark that, due to the condensation/evaporation process, the density depends on the site position $i$. 
In fact it shows a Friedel like oscillations on the boundaries, see  \cite{progress}.
Note that to obtain \eqref{eq:densite}, we have simplified by the factor $Z_L=\llangle W| G_2^{L} |V\rrangle$. 
For consistency, we constructed  a representation where this factor is non-zero, see appendix.

%

We can also compute the value of the currents. There are two types of currents: the diffusion current in the bulk, and the evaporation current. 
The diffusion current from the site $i$ to $i+1$ reads
\begin{equation}
 \langle J_{i\rightarrow i+1}^{lat} \rangle = \frac{\kappa^2}{\kappa+1}\ \ 
 \frac{d\phi^{L-i-1}+bc\phi^{2L-i-2}-c\phi^{i-1}-ad\phi^{L+i-2}}{1-ab\phi^{2L-2}}.
\end{equation}
The evaporation current at sites $(i,i+1)$ has the form
\begin{equation}
 \langle J_{i,i+1}^{eva} \rangle = -\frac{\kappa}{\kappa+1}\ 
 \frac{c\phi^{i-1}+ad\phi^{L+i-2}+d\phi^{L-i-1}+bc\phi^{2L-i-2}}{1-ab\phi^{2L-2}}.
\end{equation}

We performed several other calculations, such as the variance of the lattice current, the thermodynamical limit, and a comparison with the Macroscopic Fluctuation Theory (MFT, \cite{Bertini}). For more details about the model presented in this section, see \cite{phi}.

\section{2-species TASEP with boundaries\label{sect:2TASEP}}
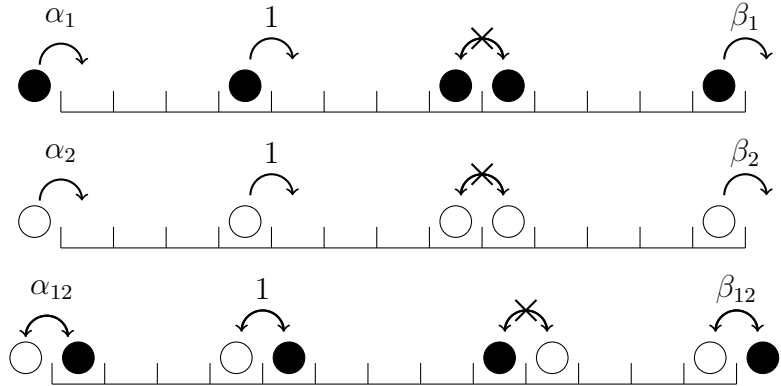
\begin{figure}[htb]
\begin{center}
\begin{tikzpicture}[scale=0.7]
\draw (-2,0) -- (11,0) ;
\foreach \i in {-2,-1,...,11}
{\draw (\i,0) -- (\i,0.4) ;}
\draw[->,thick] (-2.4,0.9) arc (180:0:0.4) ; \node at (-2.,1.8) [] {${\alpha_1}$};
\draw  (-2.5,0.5) circle (0.3) [fill,circle] {};
\draw[->,thick] (5.6,1) arc (180:0:0.4); \draw[->,thick] (6.4,1) arc (0:180:0.4); 
\draw[thick] (5.8,1.2) -- (6.2,1.6) ;\draw[thick] (5.8,1.6) -- (6.2,1.2) ;
\draw  (5.5,0.5) circle (0.3) [fill,circle] {};
\draw  (6.5,0.5) circle (0.3) [fill,circle] {};
\draw[->,thick] (1.6,1) arc (180:0:0.4); \node at (2.,1.8) [] {$1$};
\draw  (1.5,0.5) circle (0.3) [fill,circle] {};
\draw  (10.5,0.5) circle (0.3) [fill,circle] {};
\draw[->,thick] (10.6,1) arc (180:0:0.4) ; \node at (11.,1.8) [] {${\beta_1}$};
 \end{tikzpicture}
\\[1ex]
 \begin{tikzpicture}[scale=0.7]
\draw (-2,0) -- (11,0) ;
\foreach \i in {-2,-1,...,11}
{\draw (\i,0) -- (\i,0.4) ;}
\draw  (-2.5,0.5) circle (0.3) [circle] {};
\draw[->,thick] (-2.4,0.9) arc (180:0:0.4) ; \node at (-2.,1.8) [] {${\alpha_2}$};
\draw[->,thick] (5.6,1) arc (180:0:0.4); \draw[->,thick] (6.4,1) arc (0:180:0.4); 
\draw[thick] (5.8,1.2) -- (6.2,1.6) ;\draw[thick] (5.8,1.6) -- (6.2,1.2) ;
\draw  (5.5,0.5) circle (0.3) [circle] {};
\draw  (6.5,0.5) circle (0.3) [circle] {};
\draw[->,thick] (1.6,1) arc (180:0:0.4); \node at (2.,1.8) [] {$1$};
\draw  (1.5,0.5) circle (0.3) [circle] {};
\draw  (10.5,0.5) circle (0.3) [circle] {};
\draw[->,thick] (10.6,1) arc (180:0:0.4) ; \node at (11.,1.8) [] {${\beta_2}$};
 \end{tikzpicture}
\\[1ex]
 \begin{tikzpicture}[scale=0.7]
\draw (-1,0) -- (12,0) ;
\foreach \i in {-1,...,12}
{\draw (\i,0) -- (\i,0.4) ;}
\draw[->,thick] (-1.5,0.9) arc (180:0:0.4) ; \draw[->,thick] (-.7,0.9) arc (0:180:0.4) ; 
\node at (-1.,1.8) [] {${\alpha_{12}}$};
\draw  (-1.5,0.5) circle (0.3) [circle] {};
\draw  (-.5,0.5) circle (0.3) [fill,circle] {};
\draw  (2.5,0.5) circle (0.3) [circle] {};
\draw  (3.5,0.5) circle (0.3) [fill,circle] {};%
\draw[->,thick] (2.6,1) arc (180:0:0.4); \draw[->,thick] (3.4,1) arc (0:180:0.4); 
 \node at (3.,1.8) [] {$1$};
\draw  (7.5,0.5) circle (0.3) [fill,circle] {};
\draw  (8.5,0.5) circle (0.3) [circle] {};
\draw[->,thick] (7.6,1) arc (180:0:0.4); \draw[->,thick] (8.4,1) arc (0:180:0.4); 
\draw[thick] (7.8,1.2) -- (8.2,1.6) ;\draw[thick] (7.8,1.6) -- (8.2,1.2) ;
\draw  (11.5,0.5) circle (0.3) [circle] {};
\draw[->,thick] (11.6,1) arc (180:0:0.4) ; \draw[->,thick] (12.4,1) arc (0:180:0.4) ; \node at (12.,1.8) [] {${\beta_{12}}$};
\draw  (12.5,0.5) circle (0.3) [fill,circle] {};
 \end{tikzpicture}
 \end{center}
 \caption{The transition rates in the 2-species TASEP\label{fig:2tasep}}
\end{figure}

The 2-species TASEP is not a new model, it was known for a long time. It is depicted in figure \ref{fig:2tasep}.
However, for generic values of the boundary parameters $\alpha_j$ and $\beta_j$ the model is not integrable, and moreover the matrix ansatz was known only for "simple" boundaries, where one species is trapped, or when it is absent in the steady state of the model, see for instance \cite{Uchi,Arita0}. 
We classified the integrable boundaries of the model, discovering new (integrable) boundaries where both species can jump in and out of the lattice. Moreover, using the "integrable approach" we performed a matrix ansatz for these integrable boundaries.

\subsection{Integrable boundaries}
To keep this review note short, we do not show the integrable $K$-matrices, but rather give the integrable transition rates.
They are gathered in \eqref{2TASEP-left} and \eqref{2TASEP-right}, where 0 stands for holes, and 1 (resp. 2) label 
the black (resp. white) particles of figure \ref{fig:2tasep}.
The interested reader can find in \cite{open2tasep} the explicit expressions for the $K$-matrices.

\begin{eqnarray}
&\begin{array}{|c|c|c|c|}  \hline
L_1&L_2&L_3&L_4\\
\hline
\rule{0ex}{3ex}\quad 0 \,\xrightarrow{\ \mu\ }\, 1\quad&\quad 0 \,\xrightarrow{\, 1-\alpha\, }\, 1\quad&\quad 0 \,\xrightarrow{\ \alpha\ }\, 2\quad&\quad 0 \,\xrightarrow{\ \alpha\ }\, 2 \quad\\
&\quad0 \,\xrightarrow{\ \ \alpha \ \ }\, 2\quad&\quad 1 \,\xrightarrow{\ \alpha\ }\, 2\quad& \\
\quad&\quad 1 \,\xrightarrow{\ \ \alpha \ \ }\, 2\quad& &  \\
\hline
 \end{array}&\nonumber\\
 &\mb{Possible integrable rates (left boundary)}&
 \label{2TASEP-left}
\end{eqnarray}

\begin{eqnarray} 
&\begin{array}{|c|c|c|c|} 
\hline
R_1&R_2&R_3&R_4\\
\hline
\rule{0ex}{3.4ex}\quad2 \,\xrightarrow{\ \nu\ }\, 1\quad&\quad  1 \,\xrightarrow{\ \ \beta\ \ }\, 0\quad&\quad 1 \,\xrightarrow{\ \beta\ }\, 0\quad&\quad 2 \,\xrightarrow{\ \beta\ }\, 0\quad\\
&\quad 2 \,\xrightarrow{\ \ \beta\ \ }\, 0\quad&\quad 2 \,\xrightarrow{\ \beta\ }\, 0\quad& \\
\quad&\quad 2 \,\xrightarrow{\, 1-\beta\, }\, 1\quad& &  \\
\hline
 \end{array}&\nonumber\\
 &\mb{Possible integrable rates (right boundary)}&
 \label{2TASEP-right}
\end{eqnarray}

\subsection{Matrix ansatz}
Obviously, the matrix ansatz depends on which boundary conditions one chooses, and some (but not all) 
correspond to already studied cases.
 We take as an example the case of boundaries $L_2-R_3$, that was not studied up to now. 
 We introduce the following vector
\begin{equation}
 A( x)=\left(\!\begin{array}{c}
 \displaystyle
 {x^2}+G_9 x+G_8+\frac{G_7}{ x}\\[1ex]
 \displaystyle
 G_6 x+G_5+\frac{G_4}{{x}}\\[1ex]
 \displaystyle
 G_3 x+G_2+\frac{G_1}{ x}+\frac{1}{{x^2}}
 \end{array}\!\right)\quad \Rightarrow \quad A(1):
 \begin{array}{l}
  X_0 = 1+G_9+G_8+G_7 \\[1ex]
 X_1 = G_6+G_5+G_4 \\[1ex]
 X_2 = G_3+G_2+G_1+1
  \end{array}
\end{equation}
Note the expansion in the spectral parameter $x$ which is different from previous examples.
In fact, there is up-to-now no deductive way to know the expansion that one should take for $A(x)$. 
We did it by try-and-errors and brute force calculation on 'words' of 1, 2 and 3 letters: if the expansion is too restrictive (for instance stop at $x$ instead of $x^2$ in the present case), then the values of the words on the vectors $\llangle W|$ and $|V\rrangle$ will simply vanish, and we are led to take more terms in the expansion. 
The present expansion leads to an algebra with 9 generators which is much more involved than the usual (one-species) TASEP. 
However, everything needed for the matrix ansatz is still encoded by the ZF and 
GZ relations. Again, to keep this article short, we do not reproduce all the exchange relations for the 9 generators, nor the boundary relations. The interested reader can refer to the original article \cite{open2tasep}.

\subsection{Physical quantities}
We were able to compute the following quantities, that for simplicity we present in the particular case $\alpha=\frac12$ and $\beta=1$.

The partition function:
\begin{equation}
 Z_L=\llangle W|C^L |V\rrangle=(2L+1)\fc_L \fc_{L+1} \llangle W|V\rrangle,
\end{equation}
where $C=X_0+X_1+X_2$ and 
$\fc_L=\frac{1}{L+1}\left( \begin{array}{c} 2L \\L \end{array} \right)$ is the Catalan number.

The average density of  black or white particles   at site $k$:
\begin{eqnarray*}
 n_1^{(k)} &=& \frac{1}{Z_L}\llangle W|C^{k-1}\,X_1\,C^{L-k}|V\rrangle 
 =\frac{1}{\fc_{L+1}}\sum_{i=k}^{L}\frac{L-i+1}{L+2}\fc_i\fc_{L-i},  \\
 n_2^{(k)} &=& \frac{1}{Z_L}\llangle W|C^{k-1}\,X_2\,C^{L-k}|V\rrangle
 =\frac{1}{\fc_{L+1}}\sum_{i=k}^{L}\frac{i+1}{L+2}\fc_i\fc_{L-i}\;.
\end{eqnarray*}

The current $j_i$ of the particles of type $i=1,2$:
\begin{eqnarray*}
 &&j_1=\frac{1}{Z_L}\llangle W|C^{k-1}(X_1X_0-X_2X_1)C^{L-k-1}|V\rrangle= \frac{1}{2(2L+1)}, \\
 &&j_2=\frac{1}{Z_L}\llangle W|C^{k-1}X_2(X_0+X_1)C^{L-k-1}|V\rrangle= \frac{L+1}{2(2L+1)}.
\end{eqnarray*}

We also computed a representation for the matrix ansatz algebra that ensures that it is not trivial (for all values of $\alpha$ and $\beta$), i.e. which is such that 
$\llangle W|V\rrangle\neq0$, see \cite{rep-2tasep}.

\section{Conclusion and perspectives\label{conclu}}
We presented a comprehensive approach of the {matrix ansatz} in the {integrable systems} framework. Two
  key ingredients are needed: the {ZF algebra} (in the bulk) and the {GZ relations} (on the boundaries).
 In principle it allows us to construct a matrix ansatz for {\textsl{any}} integrable reaction-diffusion process (if we know the R-matrix). However, in practice the matrix ansatz algebra can be very complicated and the computation of observables can be very hard.
 
 Several points remain to be clarified. First of all, we wish to find a prescription for the {expansion} in the vector $A(x)$.
 As already mentioned, the truncation in the expansion of $A(x)$ is done case by case, on a try-and-error basis. A general (deductive) prescription would be a major step in this 'integrable matrix ansatz'.
Secondly, the construction of the representation that ensures the validity of the matrix anstz is also done case by case: a general principle to construct it would be of great help. Apart from these two points, we wish also to 
make a better use of the {spectral parameter} (with the ZF and GZ relations) to extract property of the stationary state and find a more efficient way to compute observables. Note also that there are connections with orthogonal polynomials: either in computing physical data using the fact that they are orthogonal polynomials, or to compute explicitly orthogonal polynomials starting from these physical data, see e.g. \cite{Sasamoto,time,McDo}. 
Surely, a generalisation of the matrix ansatz that would allow to get other (excited) states above the steady states is desirable. For the moment, the only way to get them is to use the Bethe ansatz.
Finally, we would like to apply our technique to
solve more complicated models: for instance we are presently working on N-species ASEP with boundaries, see \cite{N-asep} for a presentation of integrable boundaries.

\appendix

\section{Representation for the algebra of the reaction-diffusion model}
 We want to construct the two representations associated to the vectors $\llangle W |$ and $| V \rrangle$, 
 for the algebra introduced in section \ref{sect:DISSEP}:
\begin{equation}\label{alg-G}
\phi\ G_1G_2=G_2G_1 \ ;\quad
 G_1G_3=G_3G_1 \ ;\quad \phi\ G_2G_3=G_3G_2.
\end{equation}
We first introduce a realisation of this algebra, given by
\begin{equation}\label{eq:G-edA}
 G_1=e \otimes 1, \quad G_2=A(\phi) \otimes A(\phi), \quad G_3=1 \otimes d,
\end{equation}
where
\begin{equation}
 e=\sum_{n=0}^{+\infty}|n+1\rrangle \llangle n|\quad ; \quad A(\phi)=\sum_{n=0}^{+\infty}\phi^n|n\rrangle \llangle n|
 \quad ; \quad d=\sum_{n=0}^{+\infty}|n\rrangle \llangle n+1|.
\end{equation}
We have used  $\{|n\rrangle\ |\ n\geq 0\}$ as an infinite basis of the additional space, and $e$ (resp. $d$) are the lowering (resp. raising) operators. They obey 
\begin{equation}\label{eq:edA}
de=1,\quad ed=1-A(0),\quad A(\phi)e=\phi\,e\,A(\phi) \mb{and} d\,A(\phi)=\phi\,A(\phi)d.
\end{equation}
 These relations are sufficient to show that \eqref{eq:G-edA} forms a representation of the algebra \eqref{alg-G}.
 
Now, within this realization, we seek for vectors $\llangle W |$ and $| V \rrangle$ such that
\begin{equation}
 \llangle W |\big( G_1-cG_2-aG_3 \big)=0 \mb{and}
\big( G_3-bG_1-dG_2 \big) | V \rrangle =0 .
\end{equation}
A solution is given by the infinite series
\begin{eqnarray}
 |V\rrangle &=&\sum_{n,m=0}^{+\infty}v_{n,m}\,|n\rrangle \otimes |m\rrangle 
\mb{with} v_{n,m}= d^{m-n}b^n \frac{\phi^{\frac{(m-n)(m-n-1)}{2}}}{(1-\phi^2)\cdots (1-\phi^{2n})},
\\
\llangle W|&=&\sum_{n,m=0}^{+\infty} w_{n,m}\, \llangle n| \otimes \llangle m|
\mb{with}
w_{n,m}= c^{n-m}a^m \frac{\phi^{\frac{(n-m)(n-m-1)}{2}}}{(1-\phi^2)\cdots (1-\phi^{2m})}.
\end{eqnarray}
Then, one can compute that
\beano
Z_L &=& \llangle W| G_2^L |V\rrangle 
 = \sum_{n,m=0}^{+\infty} 
\frac{\big(\phi^L\,cb/{d}\big)^n\ \big(\phi^L\,da/{c}\big)^m\ \phi^{(m-n)^2}}{(1-\phi^2)\cdots (1-\phi^{2n})\times (1-\phi^2)\cdots (1-\phi^{2m})}.
\eeano

The series is convergent when 
$$
|\phi|<1\,, \quad
\left|\frac{b\,c\,\phi^L}{(1-\phi^2)d}\right|<1 \mb{and} \left|\frac{a\,d\,\phi^L}{(1-\phi^2)c}\right|<1.
$$
Thanks to the symmetry $\kappa\rightarrow -\kappa$, we can choose $|\phi|<1$.
Then, when $L$ is large enough,
the conditions are always fulfilled (when $c,d,\kappa\neq0$). 



\section*{Bibliography}
\begin{thebibliography}{10}

\bibitem{DerrReview}  B. Derrida, 
{\it Non-equilibrium steady  states: fluctuations and large deviations
of the density and of the current}, J. Stat. Mech. \textbf{0707}  (2007)  P07023.

\bibitem{BE} R.A. Blythe and M.R. Evans,
\textsl{Nonequilibrium Steady States of Matrix Product Form: A Solver's Guide,}
J. Phys. \textbf{A40} (2007) R333-R441 and \texttt{arXiv:0706.1678}.

\bibitem{Krug2}  T.  Kriecherbauer and J. Krug, {\it A pedestrian's view on 
interacting particle systems, KPZ universality and random matrices}, 
J. Phys. A: Math. Theor. {\bf 43}, 2010  403001.

\bibitem{CKZ}  
      T. Chou, K. Mallick and R. K. P. Zia,
 \textit{Non-equilibrium statistical mechanics: From a paradigmatic model to biological transport}, 
 Rep. Prog. Phys. {\bf 74}, (2011) 116601.

\bibitem{bertin}
E. Bertin, \textsl{Theoretical approaches to the statistical physics of interacting non-conservative units},
\texttt{arXiv:1608.08507}.

\bibitem{DEHP} B. Derrida, M. Evans, V. Hakim and V. Pasquier, 
\textsl{Exact solution of a 1d asymmetric
exclusion model using a matrix formulation,} 
J. Phys. \textbf{A26} (1993) 1493.

\bibitem{sklyanin} E.K. Sklyanin, 
\textsl{Boundary conditions for integrable quantum systems,} 
J. Phys. \textbf{A21} (1988) 2375.

\bibitem{ZF} A.B. Zamolodchikov and A.B. Zamolodchikov, 
\textsl{Factorized S-matrices in two
dimensions as the exact solutions of certain relativistic quantum field theory models,}
Ann. Phys. (N.Y.) \textbf{120} (1979) 253;\\
 L.D. Faddeev, 
\textsl{Quantum completely integrable models in field theory,} 
Sov. Sci. Rev. \textbf{C1} (1980) 107.

\bibitem{GZ} S. Ghoshal and A.B. Zamolodchikov, 
\textsl{Boundary S-matrix and boundary state in two-dimensional integrable quantum field
theory,} Int. J. Mod. Phys. \textbf{A9} (1994) 3841 and \texttt{arXiv:hep-th/9306002}.

\bibitem{progress}
N. Crampe, E. Ragoucy, M. Vanicat, 
\textit{Integrable approach to simple exclusion processes with boundaries. Review and progress},
\texttt{arXiv:1408.5357} and
J. Stat. Mech. \textbf{1411} (2014) P11032

\bibitem{phi}
N. Crampe, E. Ragoucy, V. Rittenberg, M. Vanicat,
\textit{An integrable dissipative exclusion process: correlation functions and physical properties},
\texttt{arXiv:1603.06796} and
Phys. Rev. \textbf{E94} (2016) 032102

\bibitem{Bertini} L. Bertini, A. De Sole,  D. Gabrielli,
G. Jona-Lasinio and C. Landim, 
 {\it  Macroscopic Fluctuation Theory for stationary non-equilibrium
 states}, J. Stat. Phys. {\bf 107}, (2002)  635.

\bibitem{Uchi} M. Uchiyama,
\textit{Two-Species Asymmetric Simple Exclusion Process with Open Boundaries,}
Chaos, Solitons \& Fractals \textbf{35}, (2008) 398 and \texttt{arXiv:cond-mat/0703660}.

\bibitem{Arita0} C. Arita, {\it Exact analysis of two-species
asymmetric exclusion process with open boundary conditions,} 
J. Phys. Soc. Jpn. {\bf 75},  (2006) 065003.

\bibitem{open2tasep}
N. Crampe, K. Mallick, E. Ragoucy, M. Vanicat, 
\textit{Open two-species exclusion processes with integrable boundaries}, 
\texttt{arXiv:1412.5939} and
J. Phys. \textbf{A48} (2015) 175002

\bibitem{rep-2tasep}
N. Crampe, M. Evans, K. Mallick, E. Ragoucy, M. Vanicat, 
\textit{Matrix product solution to a 2-species TASEP with open integrable boundaries},
\texttt{arXiv:1606.08148}

\bibitem{Sasamoto} M. Uchiyama,  T.  Sasamoto  and M. Wadati, {\it Asymmetric simple exclusion
 process with open boundaries and Ashkey-Wilson polynomials}, J. Phys. A: Math. Gen. {\bf 37}, (2004)  4985.

\bibitem{time}
N. Crampe, K. Mallick, E. Ragoucy, M. Vanicat,
\textit{Inhomogeneous discrete-time exclusion processes},
\texttt{arXiv:1506.04874} and
J. Phys. \textbf{A48} (2015) 484002

\bibitem{McDo}     L. Cantini, J. de Gier, M. Wheeler,
\textsl{Matrix product formula for Macdonald polynomials},
 J. Phys. \textbf{A48} (2015) 384001 
  and \texttt{arXiv:1505.00287}

\bibitem{N-asep}
N. Crampe, C. Finn, E. Ragoucy, M. Vanicat,
\textit{Integrable boundary conditions for multi-species ASEP},
\texttt{arXiv:1606.01018},
J. Phys. \textbf{A} (2016) to appear

\end{thebibliography}
\end{document}